\documentclass[prd,preprint,superscriptaddress,showpacs,byrevtex]{revtex4}

\usepackage{epsfig}

\newcommand{\be}{\begin{equation}}
\newcommand{\ee}{\end{equation}}
\newcommand{\bea}{\begin{eqnarray}}
\newcommand{\eea}{\end{eqnarray}}

\begin{document}

\begin{flushright}
YITP-SB-05-33
\end{flushright}

\title{Non-Perturbative Quark-Antiquark Production From a Constant 
Chromo-Electric Field via the Schwinger Mechanism }

\author{Gouranga C. Nayak } \email{nayak@insti.physics.sunysb.edu}
\affiliation{
C. N. Yang Institute for Theoretical Physics, Stony Brook University, 
SUNY, Stony Brook, NY 11794-3840, USA }

\date{\today}

\begin{abstract} 

We obtain an exact result for the non-perturbative quark 
(antiquark) production rate and its $p_T$ distribution 
from a constant SU(3) chromo-electric field $E^a$ with arbitary color 
index $a$ by directly evaluating the path integral.
Unlike the WKB tunneling result, which depends only on one gauge invariant 
quantity $|E|$, the strength of the chromo-electric
field, we find that the exact result for the $p_T$ distribution for 
quark (antiquark) production rate depends on two independent Casimir 
(gauge) invariants, $E^aE^a$ and $[d_{abc}E^aE^bE^c]^2$. 

\end{abstract} 

\maketitle
\newpage

Non-perturbative quark anti-quark pair production from a constant 
chromo-electric
field is widely employed to study hadronization at low $p_T$ in high 
energy ${\rm e}^+{\rm e}^-$ and pp collisions 
\cite{pythia}. In these approaches the color flux-tube energy density or the 
string tension is related 
to the constant chromo-electric field strength $|E|$. 
In high energy heavy-ion 
collisions at the RHIC 
and the LHC \cite{rhic} a classical chromo field
might be formed just after two nuclei pass through each other \cite{all,all1}.
In order to study the 
production of a quark-gluon plasma from a classical chromo field
it is necessary to know how quarks and gluons are formed 
from the latter. 
In a recent paper \cite{peter} we have derived a formula for the rate
for non-perturbative gluon pair production and its 
$p_T$ distribution from a 
constant SU(3) chromo-electric field with arbitary color index $a$ via
vacuum polarization. In this paper we will extend our study to 
quark-antiquark pair production.

We will not employ Schwinger's proper time method \cite{schw} 
in our calculation. 
Although this method is widely used to obtain the total fermion
production rate $dN/d^4x$ \cite{schw,yildiz1} (for a review see \cite{dune}),
this method cannot be used
to obtain the $p_T$ distribution of the rate $dN/d^4xd^2p_T$. For 
this purpose the WKB tunneling method \cite{wkb1,wkb2}
has been used in the past to approximate the $p_T$ distribution for 
the quark (antiquark) production rate.
Although the WKB tunneling result for the $p_T$ distribution 
is correct in QED it is not necessarily true in QCD because of the presence
of the non-trivial color generators in the fundamental and adjoint 
representations of SU(3). 
For this reason we will directly evaluate the path integral in this paper
and obtain an exact result for the $p_T$ distribution of
the non-perturbative quark (antiquark) production rate from a constant 
chromo-electric field with arbitary color
in the gauge group SU(3). 
We find that unlike the WKB tunneling result, which depends on one gauge 
invariant quantity $|E|$, the strength
of the chromo-electric field, \cite{wkb1,wkb2}
the exact result for the $p_T$ distribution for the 
quark (antiquark) production rate 
depends on two independent gauge invariants, $E^aE^a$ 
and $[d_{abc}E^aE^bE^c]^2$.  

We obtain the following formula for the number of non-perturbative quarks 
(antiquarks) produced per unit time, per unit volume and per unit transverse 
momentum from a given constant chromo-electric field $E^a$ 
\bea
\frac{dN_{q,\bar q}}{dt d^3x d^2p_T}~
=~-\frac{1}{4\pi^3} ~~ \sum_{j=1}^3 ~
~|g\lambda_j|~{\rm ln}[1~-~e^{-\frac{ \pi (p_T^2+m^2)}{|g\lambda_j|}}]\,,
\label{1}
\eea
where $m$ is the mass of the quark.
This result is gauge invariant because it depends on the 
following gauge invariant eigenvalues 
\bea
&&~\lambda_1~=~\sqrt{\frac{C_1}{3}}~{\cos}\theta\,,  
\nonumber \\
&&~\lambda_2~=~\sqrt{\frac{C_1}{3}}~{\cos}~ (2\pi/3-\theta)\,,  
\nonumber \\
&&~\lambda_3~=~\sqrt{\frac{C_1}{3}}~{\cos}~ (2\pi/3+\theta)\,,
\label{lm}
\eea
where $\theta$ is given by
\bea
\cos^23\theta~=3C_2/C_1^3\,.
\label{theta}
\eea
These eigenvalues only depend on two independent Casimir invariants for SU(3)
\bea
C_1~=~E^aE^a\,, ~~~~~~~~~~~
C_2~=~[d_{abc}E^aE^bE^c]^2\,,
\label{casm}
\eea
where $a,~b,~c$ ~=~1,...,8. 
Note that $0 ~\le~ {\cos}^2 3\theta ~\le~ 1$ because $C_1^3~\ge~3C_2$
and both $C_1$ and $C_2$ are positive.
The integration over $p_T$ in eq. (\ref{1}) 
reproduces Schwinger's result for total production rate $dN/d^4x$ 
\cite{yildiz1}. 

The exact result in eq. (\ref{1}) can be contrasted with the following 
formula obtained by the WKB tunneling method \cite{wkb1}
\bea
\frac{dN_{q,\bar q}}{dt d^3x d^2p_T}~=~\frac{-|gE|}{4\pi^3} ~ 
{\rm ln}[1~-~e^{-\frac{\pi (p_T^2+m^2)}{|gE|}}]\,.
\label{7}
\eea
In our result in eq. (\ref{1}) the symmetric tensor $d_{abc}$ appears. Hence 
the WKB tunneling method does not reproduce the correct result for 
the $p_T$ distribution of the quark (antiquark) production rate from a
constant chromo-electric field $E^a$. We now present a derivation
of eq. (\ref{1}). 

The Lagrangian density for a quark in a non-abelian background field
$A_{\mu}^a$ is given by
\be
{\cal{L}}~
=~\bar{\psi}^i~ [(\delta_{ij}~\hat{{p}\!\!\!\slash} 
~-~gT^a_{ij}{{A}\!\!\!\slash}^a) -m\delta_{ij}]
~\psi^j ~=~\bar{\psi}^i~ M_{ij}[A] ~\psi^j\,,
\label{laq}
\ee
where $\hat{p}_\mu$ is the momentum operator and
$T^a_{ij}$ is the generator in the fundamental representation 
of gauge group SU(3) with
$a$=1,2...8 and $i,j$~=~1,2,3. The vacuum to vacuum transition 
amplitude in the 
presence of the non-abelian background field $A_\mu^a$ is given by
\bea
 <0|0>~=~\frac{\int~[d\bar{\psi}][d\psi]~ 
e^{i\int d^4x~\bar{\psi}^j~M_{jk}[A]~\psi^k}}{
\int~[d\bar{\psi}][d\psi]~ e^{i\int d^4x~\bar{\psi}^j~M_{jk}[0]~\psi^k}} 
~ =~{\rm Det}[M[A]]/{\rm Det}[M[0]] ~=~e^{iS^{(1)}}\,.
\eea
The one loop effective action becomes
\bea
S^{(1)}_{q \bar q}~=~ -i~{\rm Tr}\,\ln
[(\delta_{ij}~\hat{{p}\!\!\!\slash} 
~-~gT^a_{ij}{{A}\!\!\!\slash}^a) -m\delta_{ij}]
+i {\rm Tr}\,\ln [\delta_{ij}~\hat{{p}\!\!\!\slash}  -m\delta_{ij}]\,.
\label{efgl}
\eea
The trace Tr contains an integration over $d^4x$,
a sum over color and Lorentz indices and a trace over Dirac matrices. 
Since the trace is invariant under transposition we also get
\bea
S^{(1)}_{q \bar q}~=~ -i~{\rm Tr}\ln
[(\delta_{ij}~\hat{{p}\!\!\!\slash} 
~-~gT^a_{ij}{{A}\!\!\!\slash}^a) +m\delta_{ij}]
+i {\rm Tr}\ln [\delta_{ij}~\hat{{p}\!\!\!\slash}  +m\delta_{ij}]\,.
\label{efg2}
\eea
Adding both the above equations we get
\bea
2S^{(1)}_{q \bar q}~=~ -i~{\rm Tr} \ln
[(\delta_{ij}~\hat{{p}\!\!\!\slash} 
~-~gT^a_{ij}{{A}\!\!\!\slash}^a)^2 -m^2\delta_{ij}]
+i{\rm Tr} \ln [\delta_{ij}~\hat{{p}}^2  -m^2\delta_{ij}]\,,~
\label{efg3}
\eea
which can be written as
\bea
2S^{(1)}_{q \bar q}~=~ -i~{\rm Tr} \ln
[(\delta_{ij}~\hat{{p}} ~-~gT^a_{ij}{{A}}^a)^2 ~+~
\frac{g}{2}\sigma_{\mu \nu}T^a_{ij}F^{a \mu \nu}
-m^2\delta_{ij}] +i {\rm Tr}\ln [\delta_{ij}~\hat{{p}}^2  -m^2\delta_{ij}]\,.~
\label{efg4}
\eea

Since it is convenient to work with the trace of the exponential
we replace the logarithm by 
\bea
\ln\frac{a}{b}~=~\int_0^\infty~\frac{ds}{s} [
e^{is~(b~+i\epsilon)}~ -e^{is~(a~+i\epsilon)}]\,.
\label{eone}
\eea
We assume that the constant electric field is along the z-axis 
(the beam direction) 
and we choose the gauge $A_0^a=0$ so that $A_3^a~=~-E^a \hat{x^0}$ \cite{peter}.
The color indices  (a=1,....8) are arbitrary. 
Since ${\Lambda}_{ij}~=~T^{a}_{ij}E^a$ has three eigenvalues we write
after diagonalization 
\bea
{(\Lambda_d)}_{ij}~=~( \lambda_1, \lambda_2, \lambda_3)\,.
\label{eigen}
\eea

The trace over the Dirac matrices (${\rm tr}_D$) give
\bea
{\rm tr}_D [e^{is\frac{g}{2}\sigma_{\mu \nu}T^a_{ij}F^{a \mu \nu}}]~=~4
\cosh(sgT^a_{ij}E^a)\,.
\eea
To reduce this problem to the motion of one harmonic oscillator, we make 
a similarity transformation \cite{peter,itz} (we also 
make a similarity transformation in the group space) and obtain
\bea
&& {\rm tr}_D~ e^{is [(\delta_{ij}~\hat{{p}} ~-~gT^a_{ij}{{A}}^a)^2 ~+~
\frac{g}{2}\sigma_{\mu \nu}T^a_{ij}F^{a \mu \nu} -m^2\delta_{ij}]} ~= 
\nonumber \\
&&~ 4~\cosh(sg({\Lambda_d})_{il})
[  e^{ip^3p_0/g(\Lambda_d)}~
e^{is({\hat{p_0}}^2 -{\hat{p_T}}^2 ~-~m^2~-~g^2{(\Lambda_d^2)} {\hat{x^0}}^2)}
~e^{-ip^3p_0/g(\Lambda_d)}~ ]_{lj}\,, 
\label{map1}
\eea
where $p_T~=~\sqrt{p_1^2+p_2^2}$ is the transverse momentum of the 
quark or antiquark (transverse to the electric field direction). 
Hence from eqs. (\ref{efg4}) and (\ref{eone}) we find
\bea
2S^{(1)}_{q\bar q}&& ~=~i~\int_0^\infty~\frac{ds}{s}~\sum_{j=1}^3~ 
{\rm tr} \Big[ 
e^{ip^3p_0/g\lambda_j}~
e^{is({\hat{p_0}}^2 -{\hat{p_T}}^2~-~m^2 
~-~g^2\lambda_j^2 {\hat{x_0}}^2 +i\epsilon)}~
e^{-ip^3p_0/g\lambda_j} \nonumber \\
&&\quad \times~[4\cosh sg\lambda_j] ~-~4e^{is({\hat{p_0}}^2 -
{\hat{p_T}}^2~-~m^2 +i\epsilon)} \Big]\,.
\label{trhg}
\eea
The trace tr denotes an integral over a complete set of $x$ eigenstates.
We add complete sets of $p_j$ eigenstates, and obtain 
\bea
2S^{(1)}_{q \bar q}~=~i~\int_0^\infty~\frac{ds}{s}~\sum_{j=1}^3
~ \frac{1}{4\pi^3}~\int d^4x~\int
d^2p_T~
e^{-is(p_T^2~+~m^2) ~-~s\epsilon}[~|g\lambda_j| 
\frac{\cosh sg\lambda_j}{\sinh s|g\lambda_j|}-\frac{1}{s}~]\,.
\label{trg}
\eea

The $s$-integral at fixed $p_T$ is convergent at s $\rightarrow$ 0, 
but the integration over $p_T$ yields an extra factor $1/s$ 
so now it seems divergent.  However charge renormalization cures this 
ultraviolet problem by subtracting also the term linear in $s$ in the 
expansion of ${\cosh sg\lambda_j}/{\sinh s|g\lambda_j|}$. 
The integral is well behaved as $s \rightarrow \infty$.
To perform the $s$ contour integration we use the well-known expansion
\be
\frac{1}{\sinh x}~=~\frac{1}{x}~+~2x~\sum_{n=1}^\infty~
\frac{(-1)^n}{x^2+n^2\pi^2}\,,
\label{sinh}
\ee
and then we formally replace $s$ by $-is$ 
(as first advocated by Schwinger in QED).
The integral is now real, except for half-circles around the poles
at $s|g\lambda_j|~=~-in\pi$ for n=1,2,3... 
The $1/x$ term in (\ref{sinh}) cancels against the $1/s$ term in (\ref{trg}).
This yields the probability for quark (antiquark) production per unit time 
and per unit volume
\bea
W_{q \bar q}~=~2{\rm Im} {{\cal L}}^{(1)}~=~\frac{1}{4\pi^3} ~\int d^2p_T~ 
\sum_{n=1}^\infty~\sum_{j=1}^3~
\frac{|g\lambda_j|}{n}~ ~e^{[-\frac{n \pi (p_T^2~+~m^2)}{|g\lambda_j|}]}\,. 
\label{gli6}
\eea
Now all that is left is to determine the eigenvalues 
$\lambda_j$ (j = 1,2,3) of the matrix 
$\Lambda_{ij}~=~T^a_{ij}E^a$ in the fundamental representation of the
gauge group SU(3). Evaluating the traces of $\Lambda_{ij}$, 
$\Lambda^2_{ij}$ and the determinant of $\Lambda_{ij}$ we find
\bea
\lambda_1~+~\lambda_2~+~\lambda_3~=~0\,,
\label{tr1}
\eea
\bea
\lambda^2_1~+~\lambda^2_2~+~\lambda^2_3~=~\frac{1}{2}E^aE^a\,,
\label{tr2}
\eea
and
\bea
\lambda_1 \lambda_2 \lambda_3 ~=~ \frac{1}{12} [d_{abc}E^aE^bE^c]\,,
\label{tr3}
\eea
the solution of which is given by eq. (\ref{lm}).

In this letter we have obtained an exact result for the rate for  
non-perturbative quark (antiquark) production and its 
$p_T$ distribution in a constant chromo-electric field $E^a$
with arbitary color index $a$ via vacuum polarization.
We have used the background field method of QCD with the gauge group SU(3). 
The $p_T$ distribution for quark (antiquark) production
can be applied at the RHIC and the LHC colliders. 
We find that, unlike the WKB tunneling method, the $p_T$ distribution of
the quark or antiquark production rate depends on two 
independent Casimir (gauge)
invariants, $E^aE^a$ and $[d_{abc}E^aE^bE^c]^2$. 

\acknowledgments
I thank Peter van Nieuwenhuizen for discussions and useful
suggestions during the completion of this work and 
Jack Smith for a careful reading of the manuscript. This work was supported 
in part by the National Science Foundation, grants PHY-0071027, 
PHY-0098527, PHY-0354776 and PHY-0345822.

\end{document}